\documentclass[structabstract]{aa}  
\usepackage{graphicx}
\usepackage{txfonts}
\begin{document}
   \title{Molecules in the disk orbiting the twin young suns of V4046 Sgr }


   \author{Joel H. Kastner
          \inst{1,2}
          \and
          B. Zuckerman\inst{3,4}
          \and
          Pierre Hily-Blant\inst{1}
          \and
          Thierry Forveille\inst{1}
          }

   \institute{Laboratoire
  d'Astrophysique de Grenoble, Universit\'e Joseph Fourier --- CNRS,
  BP 53, 38041 Grenoble Cedex, France\\
              \email{joel.kastner@obs.ujf-grenoble.fr}
         \and
         Chester F. Carlson Center for Imaging
  Science, Rochester Institute of Technology, 54 Lomb Memorial Dr.,
  Rochester, NY 14623 USA
         \and
             Dept.\ of Physics \& Astronomy, University of California, Los
  Angeles 90095 USA 
         \and 
             UCLA Center for Astrobiology, University of California, Los
  Angeles 90095 USA
             }

\titlerunning{Molecules in the V4046 Sgr disk}
\authorrunning{Kastner et al.}
   \date{Received ...; accepted ...}

 
  \abstract
  {Direct information concerning the physical
conditions and chemistry within the Jovian planet-building zones of
circumbinary disks surrounding pre-main sequence stars is essentially
nonexistent, especially for 
the more evolved pre-MS systems in which planets may already be forming or
may have formed.}
{We searched for a gaseous component within the dusty circumbinary disk
  around the nearby ($D \sim70$ pc), 12 Myr-old system V4046
  Sgr --- a tight (9 $R_\odot$ separation), short-period ($P
  \sim2.42$ day) binary with nearly equal component masses of
  $\sim$0.9 $M_\odot$ --- so as to assess the mass, chemistry, and
  kinematics of this gaseous disk.}
  {We conducted a mm-wave molecular line survey of V4046 Sgr with the 30
    m telescope of the Institut de Radio Astronomie Millimetrique
    (IRAM). We use these data to investigate the kinematics, gas mass,
    and chemical constituents of the V4046 Sgr disk.}
  {We detected rotational transitions of $^{12}$CO $^{13}$CO, HCN, CN,
    and HCO$^+$. The double-peaked CO line profiles of V4046 Sgr are
    well fit by a model invoking a Keplerian disk with outer radius of
    $\sim250$ AU that is viewed at an inclination $i=35^\circ$. We
    infer minimum disk gas and dust masses of $\sim13$ and $\sim20$
    Earth masses from the V4046 Sgr CO line and submm continuum
    fluxes, respectively.  The actual disk gas mass could be much
    larger if the gas-phase CO is highly depleted and/or $^{13}$CO is
    very optically thick.  }
  {The overall similarity of the circumbinary disk of V4046 Sgr to the
    disk orbiting the single, $\sim8$ Myr-old star TW Hya --- a
    star/disk system often regarded as representative of the early
    solar nebula --- indicates that gas giant planets are likely
    commonplace among close binary star systems. Given the relatively
    advanced age and proximity of V4046 Sgr, these results provide
    strong motivation for future high-resolution imaging designed to
    ascertain whether a planetary system now orbits its twin suns.}


   \maketitle
%

\section{Introduction}

Because precision radial velocity searches for planets have thus far
avoided close (separation $<1$ AU) binary star systems, our present
knowledge of the potential existence of planets orbiting such systems
is limited to theoretical models of giant planet formation in
circumbinary disks (Pierens \& Nelson 2008). Such models are motivated
primarily by detections of disks orbiting a handful of very young (age
$\stackrel{<}{\sim}1$ Myr) spectroscopic binaries (the best-studied
case being GG Tau; Dutrey et al.\ 1994). However, observational
constraints on the physical conditions and chemistries within the
Jovian planet-building zones of circumbinary disks --- particularly
those disks orbiting older pre-main sequence (pre-MS) binary systems,
in which planets may already be forming or may have formed --- are
essentially nonexistent.

V4046 Sgr is a close ($\sim9$ $R_\odot$ separation), 2.42 day period,
pre-main sequence (pre-MS) binary with nearly equal component masses
of $\sim$0.9 $M_\odot$ that is found far from any dark cloud (Quast et
al.\ 2000; Stempels \& Gaum 2004). Its likely membership in the
$\beta$ Pic Moving Group, which has a ``trace-back'' age of $\sim12$
Myr (Ortega et al.\ 2002; Song et al.\ 2003), implies a distance from
Earth of $\sim72$ pc (Torres et al.\ 2006, 2008).  The system has
substantial mid- to far-IR excess emission indicative of a
``transitional'' circumstellar disk, i.e., a dusty circumbinary disk
with a central gap (Jenson \& Mathieu 1997). V4046 Sgr is also a
luminous X-ray source and, furthermore, its X-ray spectrum reveals
evidence for low-temperature, high-density plasma indicative of
accretion shocks (Guenther et al.\ 2006).

In all of these respects --- save an important one, its binarity ---
V4046 Sgr closely resembles the nearby, classical T Tauri star TW Hya
(e.g., Kastner et al.\ 1997; 2002; and references therein). As a
consquence of its relatively advanced age ($\sim8$ Myr; Song et al.\
2003) and proximity to Earth ($D=53.7$ pc; van Leeuwen 2007), the TW Hya
star-disk system has been closely scrutinized and, in particular, has
served as a prototype for the study of gaseous (molecular) disks
around pre-main sequence stars via mm-wave molecular spectroscopy and
imaging (Kastner et al.\ 1997; Thi et al.\ 2004; Qi et al.\ 2004,
2006, 2008). Despite the many parallels between TW Hya and V4046 Sgr,
however, no molecular line observations of the latter system have been
reported. Hence, to ascertain whether the circumbinary disk orbiting
V4046 Sgr possesses a detectable gaseous component, and whether the
physical conditions within such a gaseous disk might be conducive to
the formation of giant protoplanets, we conducted a molecular line
survey of V4046 Sgr with the 30 m telescope of the Institut de Radio
Astronomie Millimetrique (IRAM).

\section{Observations}

\subsection{Data Acquisition and Reduction}

We observed V4046 Sgr  (RA = 18:14:10.466, dec = $-32$:47:34.50)
with the IRAM 30 m telescope (in remote mode) on 26 and 27 May
2008. The molecules and transitions observed are shown in Table 1 and
Figures 1 and 2. We observed simultaneously in either the 100 GHz (3
mm) and 230 GHz (1 mm) or 150 GHz (2 mm) and 270 GHz (1 mm) bands, and
in both polarizations in each band, using receiver combinations
A100+B100 and C230+D230 or A150+B150 and C270+D270 (all in SSB mode),
respectively. For the spectral line backends, we used both the VESPA
autocorrelator with 40 kHz and 80 kHz MHz resolution and the 1 MHz
filter banks. The weather was excellent to good ($\tau_{225} \sim$
0.05--0.3) throughout the period; time-averaged system temperatures in
both the 3 mm and 1 mm bands were in the range 200--1500 K for this
very low elevation ($\sim15-20^\circ$) source. We checked pointing and
focus (using nearby quasars as references) every 1-2 hours. Typical
pointing errors were $\sim3''$, i.e., $\sim$1/8 beamwidth for 3 mm
(FWHP beamwidth $21''$) and $\sim$1/4 beamwidth for 1 mm (FWHP
beamwidth $12''$).  Spectra were acquired in wobbler switching mode,
resulting in flat spectral baselines.

We used the CLASS\footnote{See http://iram.fr/IRAMFR/GILDAS/} radio
spectral line data 
reduction package to sum all individual spectral scans obtained in
both polarizations for a given transition, and then to subtract a
linear-fit baseline from each of these integrated spectra, calculating
channel-to-channel noise levels in the process. 
All antenna temperature measurements reported in Table 1 (i.e., peak
main-beam brightness temperature $T_{mb}$ and integrated line
intensity $I$; see \S 3.1) have been corrected assuming the beam
efficiencies ($B_{eff}$) listed in the Table.

\subsection{Results}

\begin{figure}
\includegraphics[width=9cm,angle=0]{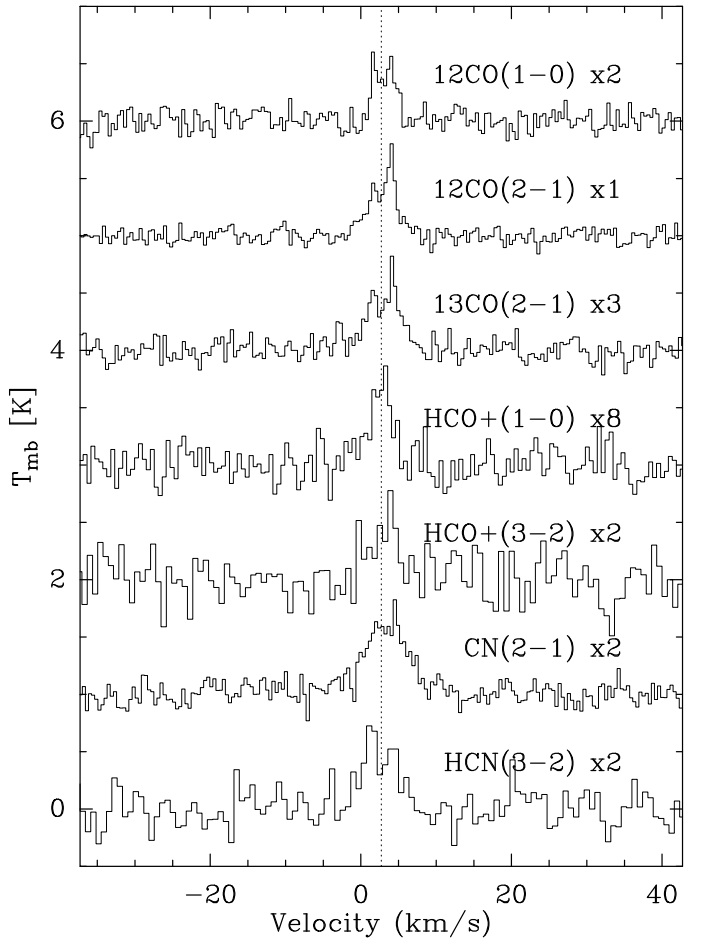}
\caption{Radio (mm-wave) molecular spectra of V4046 Sgr. Ordinate is
  velocity with respect to the 
  Local Standard of Rest (LSR) and abscissa is main beam brightness
  temperature T$_{mb}$. 
  Spectral baselines are offset 
  in T for clarity. The dashed vertical line at $+2.7$ km s$^{-1}$
  indicates the
  best-fit, line-center velocity of the CO lines.  With this radial velocity 
  and a distance from Earth of 72 pc, we calculate a Galactic space motion 
  with respect to the Sun of $U=-8.0$, $V=-16.7$, $W=-7.1$ km s$^{-1}$.}
\label{fig:specFig}
\end{figure}

\begin{table*}
\begin{center}
\caption{Molecular emission lines measured toward V4046 Sgr}
\renewcommand{\footnoterule}{}  
\begin{tabular}{cccccccc}
\hline \hline
Transition & $\nu$ & $B_{eff}$ & $T_{mb}$ & $I$ & $v_d$ & $q$  & $p_d$ \\
 & (MHz) &  & (mK) & (K km s$^{-1}$) & (km s$^{-1}$) & & \\
 (1) & (2) & (3) & (4) & (5) & (6) & (7) & (8) \\
\hline
$^{12}$CO (1--0) & 115270.204 & 0.74& 398 (56) & 1.09 (0.03) & 1.14 (0.05) & 0.56 (0.07) & 0.35 (0.04) \\  
$^{12}$CO (2--1) & 230537.990 & 0.52 & 887 (56) & 2.20 (0.03) & 1.41 (0.04) &  0.79 (0.05) & 0.24 (0.02) \\ 
$^{13}$CO (2--1) & 220398.686 & 0.55 & 346 (30) & 1.15 (0.02) & 1.6 (0.1) & 0.80 (0.08) & 0.26 (0.04) \\  
HCO$^+$ (1--0) & 89188.523 & 0.77 & 91 (27) & 0.32 (0.04) & 6.3* (1.3) & ... & ... \\  
HCO$^+$ (3--2) & 267557.625 & 0.45 & 251 (130) & 1.19 (0.26) & 4.7* (1.0) & ... & ... \\  
CN (2--1) & 226875.000 & 0.53 & 314 (42) & 2.24 (0.09) & 6.7* (0.4) & ... & ... \\
HCN (3--2) & 265886.432 & 0.45 & 225 (26) & 1.30 (0.15) & 5.4* (0.8) & ... & ... \\
H$_2$CO (2$_{21}$--1$_{11}$) & 140839.000 & 0.70 & $<$40  & $<$0.2 & ... & ... & ... \\
\hline
\end{tabular}
\end{center}

\footnotesize{NOTES --- For CO lines, peak main-beam brightness
  temperature ($T_{mb}$) and integrated line intensity ($I$) obtained
  empirically; one-half the difference between the red and blue peak
  velocities ($v_d$), radial temperature power law index ($q$), and
  outer disk cutoff parameter ($p_d$) obtained from fits to disk model
  line profiles. For other lines, $T_{B,max}$ and $I$ obtained from
  Gaussian fit to line profile. For these (Gaussian fit) lines
    the value of $v_d$ (flagged by an asterisk) represents the FWHM
  of the best-fit Gaussian. Numbers in parentheses indicate formal
  (1$\sigma$) uncertainties in best-fit parameter values. }

\end{table*}


We detected lines of $^{12}$CO $^{13}$CO, HCN, CN, and HCO$^+$ from
the circumbinary disk of V4046 Sgr. Measured line parameters are
reported in Table 1. For the CO lines --- whose profiles are sharply
double-peaked (with velocity separation $v_d$ between the peaks) ---
peak and integrated line intensities ($T_{mb}$ and $I$, respectively)
were determined empirically (from the peak and velocity-integrated
main beam brightness temperatures, respectively), while $v_d$ was
determined from a parametric fit of a model of molecular emission from
a Keplerian disk (see \S 3.1). It appears that the HCN and CN
  lines, as well as the HCO$^+$(3--2) line, also display double peaks
  at the approximate velocities of those detected in the CO lines. In
  the case of CN and HCN, the line profiles include contributions from
  closely-space hyperfine components (e.g., Skatrud et al.\ 1983),
  which tends to ``wash out'' their potential double-peaked line
  shapes. Hence, simple single-Gaussian fits served to determine these
  line parameters (including HCO$^+$(3--2), for which the
  signal-to-noise ratio renders a two-Gaussian fit unreliable). 

Substantial asymmetries are present in the two CO(2--1) line profiles
(and, possibly, HCO$^+$(3--2)), wherein the red peak is noticeably
stronger than that of the blue. This contrast, which is not apparent
in the $^{12}$CO(1--0), CN, or HCN lines, appears more profound than
the slight difference between the red and blue peak intensities
typical of pre-MS disk CO lines (see, e.g., Fig.\ 6 in Thi et al.\
2004). We consider the potential origins for these CO(2--1) line
profile asymmetries in \S 3.1.

\section{Analysis and Discussion}

\subsection{Disk structure and  kinematics}

\begin{figure}
\includegraphics[scale=0.5,angle=0]{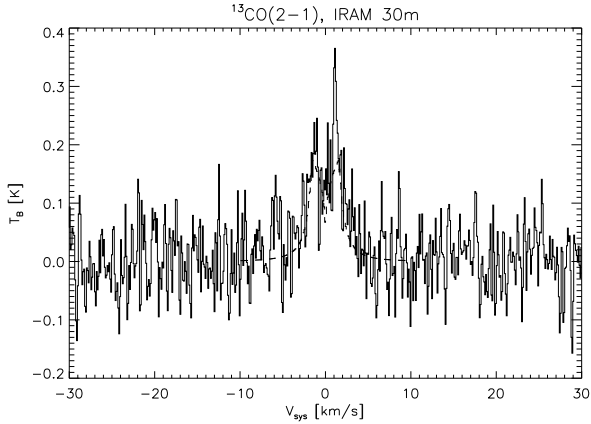}
\caption{$^{12}$CO(2--1) line profile of V4046 Sgr (solid line) overlaid with
  best-fit Keplerian disk model profiles (dashed line). Ordinate is
  velocity with respect to the systemic velocity of V4046 Sgr
  determined from the model fitting
  ($V_{LSR}=2.7$ km s$^{-1}$). }
\label{fig:COprofiles}
\end{figure}

To ascertain the kinematic characteristics of the V4046 Sgr disk, we
fit the CO line profiles using the Kastner et al.\ (2008)
parameterization of the detailed Keplerian disk model presented in
Beckwith \& Sargent (1993). This model fitting exercise directly
yields an estimate of outer disk rotation velocity ($v_d$, equivalent
to the half-value of the velocity separation of the red and blue peaks
in the line profiles), as well as indications of the slope ($q$) of
the disk radial temperature profile $T \propto r^{-q}$ (obtained
  from the shape of the high-$v$ line wings) and the slope ($p_d$) of
  the inner line profile (i.e., between $v=v_d$ and $v=0$). In the
  case of an edge-on disk, $p_d$ provides a measure of the definition
  of the disk outer edge (i.e., $p_d=1$ corresponds to a sharp outer
  edge and values $p_d<1$ indicate lack of a sharp edge; Kastner et
  al.\ 2008). However, in the present case of a disk that is likely
  viewed at intermediate inclination (see below), the central regions
  of the CO line profiles can be ``filled in'' by contributions from
  gas spanning a wide range of disk radii (e.g., Fig.\ 3 in Beckwith
  \& Sargent 1993), such that $p_d$ is less straightforward to
  interpret.

  Results obtained from these parametric fits to the CO line profiles
  are listed in Table 1.  In Fig.\ 2 we display the best-fit model
$^{12}$CO(2--1) line profile, overlaid on the IRAM spectrum. This
comparison of observed and model line profiles highlights the CO(2--1)
line profile asymmetries mentioned in \S 2.2, as such asymmetries are
not predicted by standard Keplerian disk models. Two potential
explanations for these line profile asymmetries present
themselves. First, it is possible that the disk structure is
intrinsically asymmetric. Because the similarity of the
$^{12}$CO(2--1) and $^{13}$CO(2--1) profiles appears to rule out
optical depth effects as responsible for the line profile asymmetries
--- and it is unlikely that an azimuthal temperature gradient, if
present, would be large enough to produce a measurable contrast in the
red and blue peak intensities --- the intrinsic asymmetry would most
likely result from unequal beam filling factors for the redshifted and
blueshifted emission. This in turn might suggest that, e.g., the
redshifted disk has a larger opening angle than that of the
blueshifted disk. Such an asymmetry could be caused by a perturbing
body, such as a planet or low-mass companion star. The relative
symmetry of the $^{12}$CO(1--0) line profile would then also imply
that such an intrinsic disk asymmetry is temperature-dependent.

A second, arguably more plausible, possibility is that the CO(2--1)
source is marginally resolved by the 30 m and that the telescope was
slightly mispointed.  Specifically, the measured pointing errors of
$\sim3''$ (\S 2.1) are similar to the apparent disk radius inferred
from the CO line profile analysis (see below). If the telescope
beam were centered on the peak of redshifted disk emission --- resulting
in a $\sim1/2$ beamwidth pointing offset for the blueshifted emission ---
the observed factor of $\sim2$ contrast between red and blue peak
intensities would result. Such a systematic mispointing
furthermore would be consistent with the relative symmetry of the
$^{12}$CO(1--0) line profile since, at this frequency, the source
should be unresolved and $\sim3''$ pointing errors should be
unimportant. Direct (interferometric) imaging of the molecular disk of
V4046 Sgr is required to distinguish between these alternative
(intrinsic disk asymmetry vs.\ pointing error) explanations for the
CO(2--1) line profile asymmetries.

Ignoring the line profile asymmetries, if one adopts the central star
masses determined by Stempels \& Gahm (2004) --- i.e., roughly equal
components of $0.9$ M$_\odot$ --- and assumes that the inclination of
the disk is identical to that of the V4046 Sgr binary ($i=35^\circ$;
Quast et al.\ 2000), then the best-fit values of $v_d \sim 1.5$ km
s$^{-1}$ obtained from the CO(2--1) line profile fitting (Table 1)
indicate a circumbinary disk outer radius of $\sim250$ AU. The smaller
best-fit $v_d \sim 1.1$ km s$^{-1}$ for the CO(1--0) line is
indicative of a somewhat larger effective outer disk radius in this
transition, consistent with its tracing colder gas at lower critical
density in the outermost regions of the disk.  The
best-fit values of $q$ are compatible with the ``canonical'' range of
disk temperature gradients obtained via ``standard'' disk models
(i.e., between $q=1/2$ and $q=3/4$; e.g., Omodaka et al.\ 1992;
Beckwith \& Sargent 1993).

An upper limit on the disk inner radius can be roughly imposed by the
extent of the line wings detected in the CO line profiles. These wings
are seen to extend to at least $\pm 5$ km s$^{-1}$, suggesting an
inner radius not larger than $\sim20$ AU for a central mass of 1.8
$M_\odot$ and $i=35^\circ$. Indeed, the presence of $\sim200$ K dust
in the disk (\S 3.2) indicates that the molecular disk may extend to
within $\sim3$ AU of the star.

\subsection{Disk gas and dust masses}

To estimate a lower limit on the disk gas mass from the CO line
intensities, we follow the method described in Kastner et al.\ (2008),
Zuckerman et al. (2008), and references therein.  We adopt $D = 72$ pc
and a CO:H$_2$ number ratio of $10^{-4}$, and assume that the mean gas
temperature in the region of the disk from which the CO lines
originate is similar to that of the dust ($\sim40$ K; see below). From
the measured $^{12}$CO(2--1)/$^{13}$CO(2--1) line ratio ($\sim 2$;
Table 1), we infer that the $^{12}$CO optical depth is at least
$\tau_{\rm ^{12}CO}\sim 45$, where this minimum value 
is based on the assumptions that $^{12}$C/$^{13}$C $= 89$ by number
and that $^{13}$CO is optically thin. The $^{12}$CO line intensity
then indicates that at least $4\times10^{-5}$ $M_\odot$ ($\sim$13
Earth masses) of gas resides in the circumbinary disk orbiting V4046
Sgr.  

We emphasize that this CO-based estimate of the V4046 Sgr disk gas
mass is likely just a stringent lower limit, because CO may be
significantly depleted relative to H$_2$ due to
photodissociation or freezeout onto dust grains (such that CO:H$_2$
$<<10^{-4}$), and/or $^{13}$CO may be optically thick.  Regarding the
latter possibility, the $^{12}$CO/$^{13}$CO line ratio measured for
V4046 Sgr is similar to that of the pre-MS molecular disk model
described in Omodaka et al.\ (1992), who assumed a disk mass two
orders of magnitude larger than the minimum mass just obtained for
V4046 Sgr. Furthermore, the best-fit values of $q$ for the CO(2--1)
lines (Table 1) are consistent with a disk temperature gradient of $T
\propto r^{-3/4}$, characteristic of optically thick $^{13}$CO
emission (Beckwith \& Sargent 1993). Hence, the assumption $\tau_{\rm
  ^{13}CO} < 1.0$ likely does not hold, implying the disk gas mass may
be significantly larger than $\sim$13 Earth masses. Observations of
C$^{18}$O would help establish $\tau_{\rm ^{13}CO}$ and thereby better
constrain the mass of molecular gas orbiting V4046 Sgr.

  Regardless, the V4046 Sgr disk gass mass estimated above is a factor
  $\sim2$ larger than the gas mass orbiting TW Hya (after correcting
  for the revised distance of 45 pc) as estimated by Kastner et al.\
  (1997) under the same fundamental assumptions, i.e., that CO:H$_2$
  $=10^{-4}$ and $^{12}$C/$^{13}$C $= 89$ by number and
  $\tau_{\rm ^{13}CO} < 1.0$. Morever, the $^{12}$CO(2--1)/$^{13}$CO(2--1)
  line ratio measured for TW Hya ($\sim7$; Kastner et al.\ 1997)
  suggests that $\tau_{\rm ^{13}CO} < 1.0$ is a better approximation in the case
  of TW Hya. Thus it appears that the
  V4046 Sgr disk likely harbors a molecular gas mass larger than that
  of the disk orbiting TW Hya.

Given an estimate of the dust temperature, the disk dust mass (M$_d$)
can be estimated in the standard way (e.g., equation [3] in Zuckerman
2001) from the submillimeter flux of V4046 Sgr (770 mJy at 800 $\mu$m;
Jensen et al.\ 1996). A spectral energy distribution (SED) comprised
of IRAS and submm (Jensen et al.\ 1996) fluxes indicates a temperature
of $\sim40$ K for the bulk of the dust mass, although it is clear from
the SED that dust is present with a range of temperatures up to about
200 K (J. Rhee, 2008, private comm.). Adopting a dust temperature of 40
K, then M$_d = 6\times10^{-5}$ $M_\odot$ (20 Earth masses). This
estimate is similar to that obtained on the basis of submm data
by Webb (2000) for the TW Hya disk dust mass. If the 40
K grains behave like black bodies, then the orbital semi-major axis of
the ``cold'' dust disk is $\sim$80 AU, i.e., a factor $\sim3$ smaller
than the outer disk radius obtained from the CO line profiles. This
discrepancy between the inferred radii of the dust and gas disks is
similar to that reported for many other pre-MS disks and  is most likely an artifact of the sharp
truncation of the outer disk imposed in the Keplerian disk model
(Hughes et al.\ 2008b). It is also possible that a population of colder
grains is present beyond $\sim80$ AU, however.



\subsection{HCN, CN, and HCO$^+$: comparison with other pre-MS
  star/disk systems}

\begin{figure}[htb]
\includegraphics[width=9cm,angle=0]{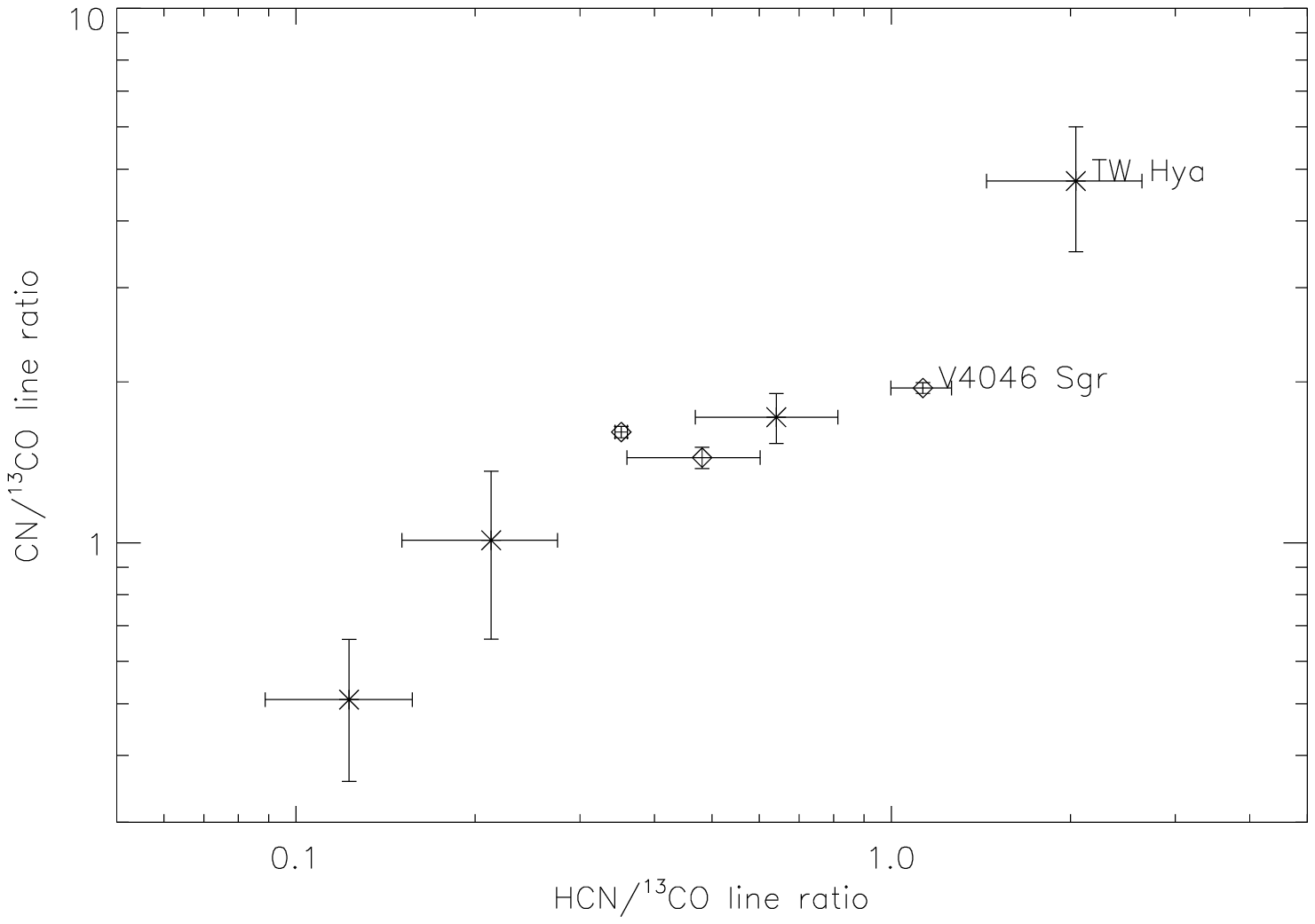}
\includegraphics[width=9cm,angle=0]{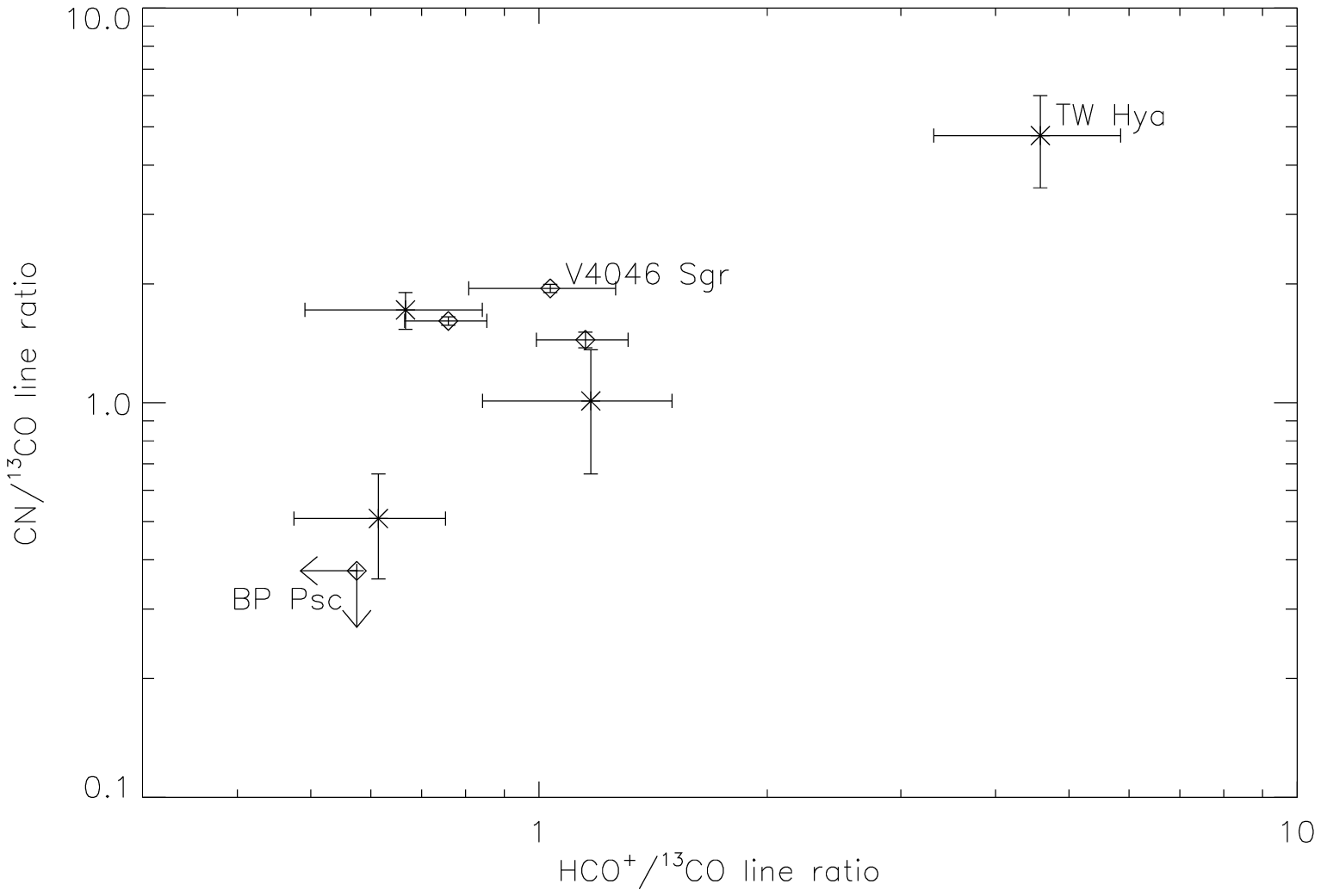}
\includegraphics[width=9cm,angle=0]{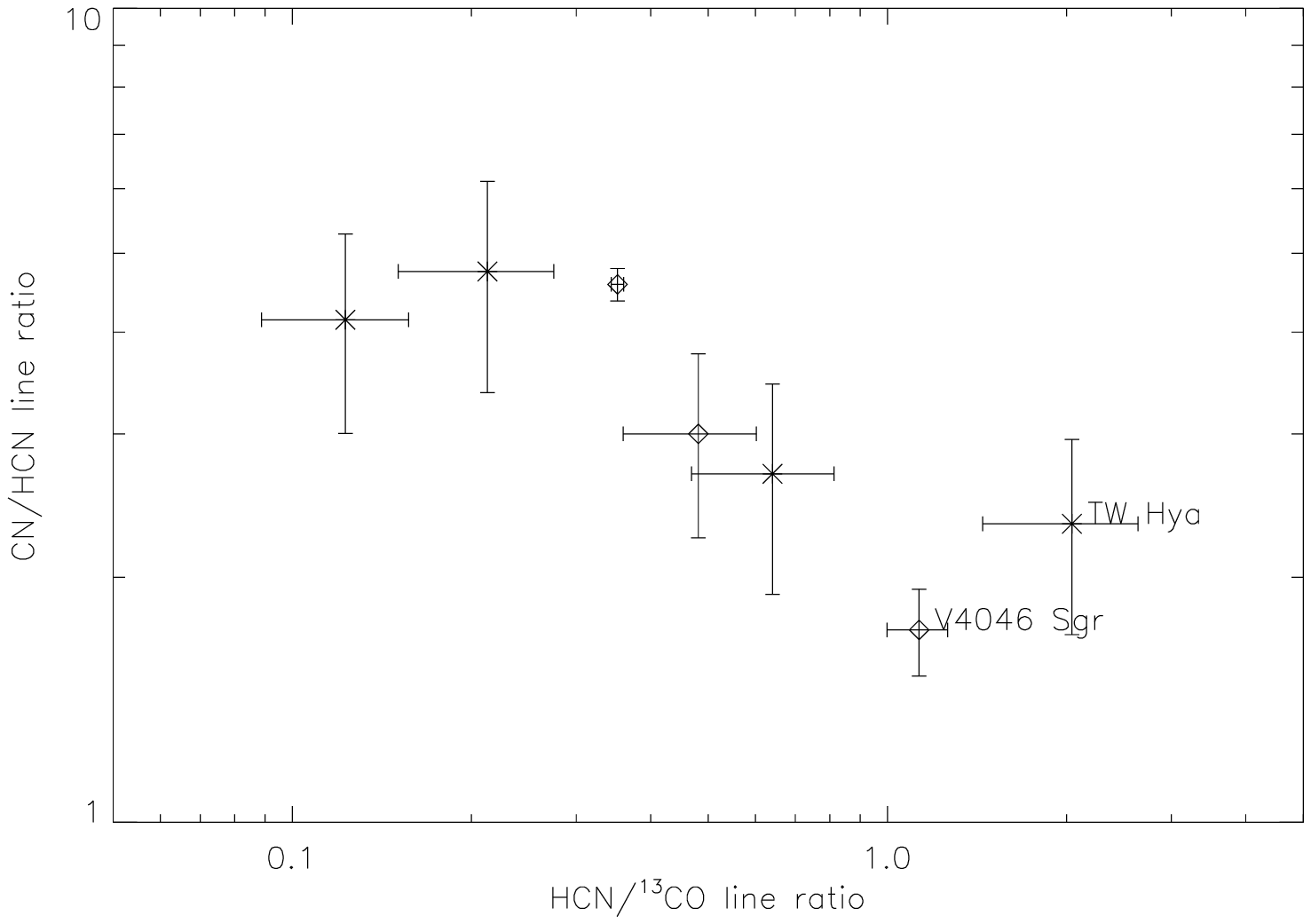}
\caption{CN/$^{13}$CO vs.\ HCN/$^{13}$CO line ratios (top),
  CN/$^{13}$CO vs.\ HCO$^+$/$^{13}$CO line ratios (middle), and CN/HCN
  vs.\ HCN/$^{13}$CO line ratios (bottom) for V4046 Sgr and for the
  (6) other pre-main sequence stars measured to date in these lines,
  including the single star TW Hya (in the middle panel, we also
  include the unusual dusty field star BP Psc, which may be a post-MS
  object; see Kastner et al.\ 2008 and references therein). Data
  points corresponding to JCMT 15 m telescope measurements of the
  integrated line intensities in the HCO$^+$(4-3), HCN(4-3), CN(3-2),
  and $^{13}$CO(3--2) transitions are indicated with ``X''; points
  corresponding to IRAM 30 m telescope measurements of the integrated
  line intensities in the HCO$^+$(3-2), HCN(3-2), CN(2-1), and
  $^{13}$CO(2--1) transitions are indicated with diamonds. }
\end{figure}

In Fig.\ 3 we plot various (sub)mm-wave molecular line ratios for
V4046 Sgr, TW Hya, and five other pre-MS molecular disk systems for
which emission in $^{13}$CO, CN, HCN, and HCO$^+$ have all been
measured (Kastner et al.\ 2008 and references therein). Note that
  the line ratio diagrams in Fig.\ 3 compare results from multiple
  transitions of these species. Hence --- while systematic errors due
  to instrumental effects may be negligible (Kastner et al.\ 2008) ---
  optical depth effects and departures from LTE (especially in the
  case of the CN lines; van Zadelhoff et al. 2003) represent potential
  complications when attempting to compare the line ratios measured
  for the various disks. That caveat aside Fig.\ 3 indicates that
V4046 Sgr lies at the upper range of pre-MS star/disk systems in terms
of its CN and HCN emission line strengths relative to $^{13}$CO, with
only TW Hya displaying larger HCN/$^{13}$CO and CN/$^{13}$CO
ratios. However, TW Hya still stands alone as a much stronger HCO$^+$
source (relative to $^{13}$CO) in comparison to V4046 Sgr and all
other pre-MS star/disk systems.
 
The large relative intensities of emission lines of CN and HCO$^+$,
accompanying strong HCN emission, indicate that high-energy (UV and
X-ray) radiation originating with the central stars and/or star-disk
interactions has significantly and similarly impacted the chemistries
of the V4046 Sgr and TW Hya disk systems (Lepp \& Dalgarno 1996; Thi
et al.\ 2004; Kastner et al.\ 2008 and references therein). Models of
X-ray ionization of molecular gas predict monotonically increasing CN,
HCN, and HCO$^+$ abundances at constant CO abundance as X-ray
ionization rate increases (see Figs.\ 2,3 in Lepp \& Dalgarno),
whereas increasing UV photodissociation rate tends to drive up the
CN/HCN ratio (e.g., van Zadelhoff et al. 2003; Boger \&
Sternberg 2005). It is therefore intriguing that the CN/HCN ratios
measured for the two $\sim10$ Myr-old classical T Tauri stars are
similar to those of much younger pre-MS stars even though their
HCN/$^{13}$CO and CN/$^{13}$CO line ratios are significantly larger
than those of the other pre-MS stars (by factors of up to $\sim$10 for
V4046 Sgr and $\sim20$ for TW Hya; Fig.\ 3). Given these results,
  and the fact that the X-ray luminosities of TW Hya and V4046 Sgr
are at least a factor $\sim3-10$ larger than those of the other pre-MS
stars plotted in Fig.\ 3 (see Kastner et al.\ 2002, 2008; Guenther et
al.\ 2006), one might speculate that X-ray ionization --- as
  opposed to UV photodissociation --- is driving the circumstellar
  disk chemistry in these two older TTS star-disk systems. However, it
  is clearly premature to conclude as much, in advance of detailed
  radiative transfer calculations (along the lines of those described
  in, e.g., van Zadelhoff et al. 2003) for all of the disks included
  in Fig.\ 3. Indeed, given a model in which X-ray ionization
  dominates over UV photodissociation in both disks, one must explain
why the HCO$^+$/$^{13}$CO ratio of V4046 Sgr is much smaller than that
of TW Hya, and is instead similar to those measured for younger pre-MS
systems.

\section{Conclusions}

The detections of circumstellar CO, HCN, CN and HCO$^+$ presented here
demonstrate that the close binary star V4046 Sgr is orbited by a rich
molecular disk.  At an age of 12 Myr, V4046 Sgr may be the oldest
known classical T Tauri star and, at
a distance from Earth of only $\sim$72 pc, it is also the second
closest (after TW Hya).  The only other star comparably close to Earth
as TW Hya and V4046 Sgr and known to possess a substantial orbiting
molecular disk is the A-type star 49 Cet (Zuckerman et al. 1995;
Hughes et al 2008a).  Given the many young stars known to reside
within 100 pc of Earth (Torres et al. 2008), it remains a mystery why
these 3 stars, which span a spectral type range from K7 to A1, should
have been able to retain so much orbiting molecular gas for so long
(49 Cet is of uncertain age, but it may be older than V4046 Sgr).

The similarity of the composition of the molecular circumbinary disk
orbiting the $\sim12$ Myr-old, close binary V4046
Sgr to that of the molecular disk orbiting the single, $\sim8$ Myr-old
star TW Hya (\S 3.3) --- and the likelihood that the disk gas mass of V4046 Sgr
is, if anything, even larger than that of TW Hya (\S 3.2) ---
indicates that the early evolution of circumbinary disks can proceed
in a manner closely analogous to that of single-star
disks. Furthermore, if the TW Hya disk is indeed an appropriate system
to serve as analog to the solar nebula just at or after the epoch of
Jovian planet formation, then the many parallels between V4046 Sgr and
TW Hya would indicate that gas giant planets are commonplace among
close binary star systems. Given the age and proximity of V4046 Sgr,
the results presented here therefore provide strong motivation for
future high-resolution imaging designed to ascertain whether a
planetary system now orbits its twin suns.

\acknowledgements{We thank Joseph Rhee for his analysis of the SED of
  V4046 Sgr and the anonymous referee for useful comments.
  J.H.K. thanks the staff of the Laboratoire d'Astrophysique de
  Grenoble for their support and hospitality during his yearlong
  sabbatical visit to that institution.  This research was partially
  supported by a NASA grant to UCLA.}

\end{document}